# Optimization of Balanced Detector for Coherent Receiver on Generic InP Platform by PSO

Dhiman Nag, Weiming Yao, and Jos J. G. M. van der Tol

*Abstract*—Balanced photodetector (BPD) is an important component for high-speed coherent receiver. Optimization strategy of waveguide-based multi-quantum well (MQW) BPDs, operating at 1550 nm is demonstrated on generic InP platform. Design parameters of BPD are optimized towards achieving the highest bandwidth for a responsivity through an algorithm based on Particle Swarm Optimization (PSO). We do so by establishing an equivalent circuit model of BPD and analyzing its opto-electronic transfer function through numerical modelling. We address the major bottlenecks of high-speed BPDs: transit time of generated carriers and RC loading in our model. The algorithm is able to provide multiple combinations of design parameters with the same output characteristics. Design methodology to integrate laser with optimized BPD is presented to successfully implement coherent receiver.

*Index Terms*—photonic integrated circuits, high-speed coherent receiver, Particle Swarm Optimization, MATLAB

## I. INTRODUCTION

Coherent balanced receivers are commonly used in high-speed photonic communication links operating at 1550 nm owing to their ability to suppress laser relative intensity noise (RIN) and to achieve high common mode rejection ratio (CMRR), high noise suppression etc. [1]–[3]. They also enable the differential phase shift keying (DPSK) schemes due to balanced detection, which has nearly 3-dB more sensitivity compared to OOK [4], [5]. For successful implementation of balanced receivers, balanced detectors (BPDs) play a major role [6]. A BPD consists of two reverse-biased near-identical and near-isolated photodetectors (PDs) [6]. Separate Optical inputs are fed to these PDs and their difference current is taken as the output. Different material platforms are investigated by many groups for realizing high-speed detectors at 1550 nm [7]–[9]. However, a generic integration platform is essential to embed several active and passive photonic building blocks (BBs) on the same chip that enables realizing broad range of applications including telecommunication, sensing, quantum key distribution, imaging etc. [10]. This work is focused on optimization of balanced detector (BPD) to be implemented in a coherent balanced receiver on a generic integration platform, based on InP/InGaAsP material system using monolithic butt-coupling technology [10]–[12]. The active components have quaternary InGaAsP core, tuned for emission or absorption. And, the passive components have a core of ideally non-absorbing quaternary material. In order to exploit the benefits of generic platform for coherent balanced receiver, BPD needs to be designed for optimum performance and monolithically integrated with laser, which works as the local oscillator. Multiple regrowth steps during butt-joint technology greatly increase fabrication complexity and process time that leads to high cost and low yield. MQW stack is used for laser to enhance its emission efficiency [13]. We consider the same MQW stack for BPD to avoid additional regrowth. Two most essential figures of merit of detectors are its bandwidth and responsivity. They possess different and often opposite dependencies on certain dimensional parameters [14]. This is attributed to the fact that high responsivity requires active region of PDs to be thick and long, which leads to lower bandwidth owing to high carrier transit time and high RC component, respectively. Since BPD is a part of coherent receiver, the later shows similar trade-off among its output current and bandwidth. Travelling wave photodetectors (TW-PDs) can in principle achieve both the goals: high bandwidth and high responsivity [14]. But long TW-PDs lead to high RF losses [15]. Also, they are designed for a narrow wavelength range to attain velocity matching between optical and RF signals. Since we intend to develop the generic InP platform for a wide wavelength range, we have focused on lumped BPD in our work which can be integrated with a widely tunable laser. Several research groups have tried to achieve detectors with bandwidth more than 100 GHz by engineering the layerstack [16]–[18]. But these approaches focus on stand-alone PDs and overlook generic integration. We target to optimize BPDs on generic InP platform towards achieving the highest bandwidth for a particular responsivity, which depend on physical dimensions and layerstack [19], [20]. It is done by creating an equivalent circuit model of BPD and transforming various physical parameters to the corresponding circuit elements. Equivalent circuits are widely used in the electrical engineering as they are more intuitive, easy to process and moderate in computation [21]. Parameters, that are specific to the generic platform are not tuned for BPD optimization e.g. number and thickness of quantum wells, material composition of quantum wells and barriers, wavelength of the incoming

Manuscript received xx Xxx 2023; Date of publication xx Xxx 2023. This work is funded by the H2020 ICT PICaboo project (contract No. 101017114) under the photonics PPP. *(Corresponding author: Dhiman Nag).*
Dhiman Nag is with the Photonic integration group, Eindhoven University of Technology, the Netherlands (e-mail: d.nag@tue.nl).
Weiming Yao is with the Photonic integration group, Eindhoven University of Technology, the Netherlands (e-mail: w.yao@tue.nl).
Jos J. G. M. van der Tol is with the Photonic integration group, Eindhoven University of Technology, the Netherlands (e-mail: j.j.g.m.v.d.tol@tue.nl).
Digital object identifier: xx.xx.xxxxx

optical signal, contact resistance, material properties of the layerstack etc. We focus on achieving highest bandwidth at a responsivity by optimizing four physical parameters: length of each detector, thickness of core, position of MQW inside core, and barrier thickness between two QWs. These parameters impact RC component and transit time of BPD, which in turn affect its bandwidth and responsivity, as discussed in section IV.C. An Opto-electronic transfer function is derived from the equivalent circuit and is used for all calculations. Conventional optimization of multiple parameters needs a lot of time. So, we employ Particle Swarm Optimization (PSO), a metaheuristic algorithm proposed by Kennedy and Ebarhart in 1995 in an attempt to model social interaction among school of fish or flock of birds [22]. PSO can generate high-quality solutions within shorter calculation time, which has a robust convergence characteristics compared to other heuristic algorithms such as ant colony optimization, simulated annealing, firefly algorithm, genetic algorithm etc. [23]. We also present how optimized BPD can be integrated with a laser using the same layerstack to implement coherent receiver.

This article is organized as follows: section II presents the layerstack for BPD, and its device structure followed by its equivalent circuit model and the transfer function. DC and RF responsivity are presented in the section III. BPD optimization by PSO, power budget estimation of coherent receiver and strategy for integrating optimized BPD with laser are presented in section IV. Section V concludes the work.

## II. MODELLING BACKGROUND

This section starts with the cross-section and device structure of waveguide PDs on our generic InP platform. We represent the BPD with an electrical circuit by translating various physical parameters into equivalent circuit components.

### A. Device Structure of Single Detector

Waveguide-based detectors are considered for this work due to suitability of photonic integration. Besides, length of the waveguide can be designed so that majority of photons are absorbed without impacting the transit time [14], [19]. Fig. 1 shows the InP-based *p-i-n* heterostructure, used for designing the BPD. Core of waveguide, containing the absorbing MQW, is sandwiched between *p*- and *n*-doped InP layers. Quantum wells are made of quaternary InGaAsP, whose composition is chosen such that the incoming optical signal of a particular wavelength (1550 nm in our case) is absorbed and the cladding layers remain transparent to the wavelength. Position of the MQW stack inside the core is represented with p-offset and n-offset, also mentioned in the caption. Width of the PD is $W_{PD}$. Thickness of the core is denoted as $d_{act}$, which has a lower bandgap and higher refractive index, while the cladding layers have higher bandgap and lower refractive index. Such profiles in refractive index and bandgap help optical mode to be confined in the core region. Schematic of a single detector structure is shown in fig. 2(a). Polyimide is used as a polymer to planarize the topological difference between *p*-metal and *n*-metal. Generated electron-hole pair in the MQW stack are separated by the electric field, present in the intrinsic region under reverse bias and are collected in the *n*-InP and *p*-InP layers, respectively. The equivalent circuit components for different regions are indicated in fig. 2(b), which are discussed in the below sub-section.

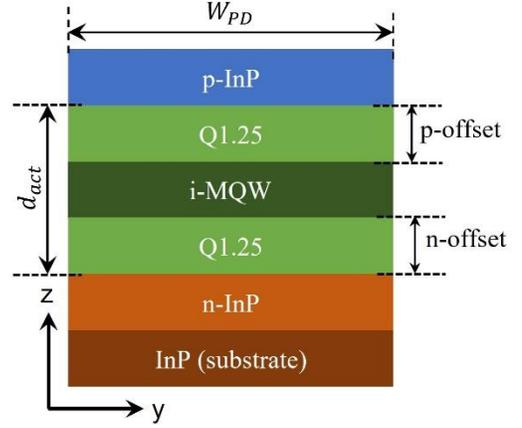

Fig. 1. Cross-section of the layerstack used for designing balanced detector. Core comprises of MQW stack sandwiched between two non-absorbing ternary layers. Distance from *p*-InP and *n*-InP to MQW are termed as p-offset and n-offset, respectively.

### B. Equivalent Circuit Model of Balanced Detector

Schematic and equivalent circuit of BPD are shown in fig. 3. In case of waveguide based BPD, optical modes with different phases (180º apart) get detected by two PDs. Difference current ($I_{out} = I_1 - I_2$) is extracted from the intersection of these PDs. $I_1$ and $I_2$ are generated AC current by the detectors $PD_1$ and $PD_2$, respectively. Phase difference between these two currents is 180º to reflect the phase difference of two optical signals onto $PD_1$ and $PD_2$. Fig. 3(b) shows the equivalent circuit model of a BPD. Active region of the PDs is represented with a junction resistance $R_j$ and an effective junction capacitance $C_{eff}(= C_j||C_{poly})$ in parallel. Junction capacitance inside the PD is $C_j$ and $C_{poly}$ is the capacitance due to polyimide in between *p*- metal and $n$-InP. $p$-metal of width 20 $\mu$m is considered atop $p$-InP centered with the waveguide. Series resistance and load resistance are $R_s$ and $R_L$, respectively. $R_s$ is determined by the contact resistance and horizontal distance between p-metal and n-metal, which is defined by the generic technology. $C_{pd}$ is the capacitance due to metal pads. The current source $I_{AC}$ denotes generated RF current during detection. Junction of the PDs can be converted with Norton equivalent circuit, as shown in fig. 3(c). Norton equivalent current $I'_{AC}$ is computed and expressed in the figure. Junction impedance $R_j$ is ignored because of very high value (10−100 M$\Omega$) under reverse bias. Fig. 3(d) shows the AC equivalent circuit of BPD by replacing the PD junction with Norton equivalent circuit and shorting the DC bias source.

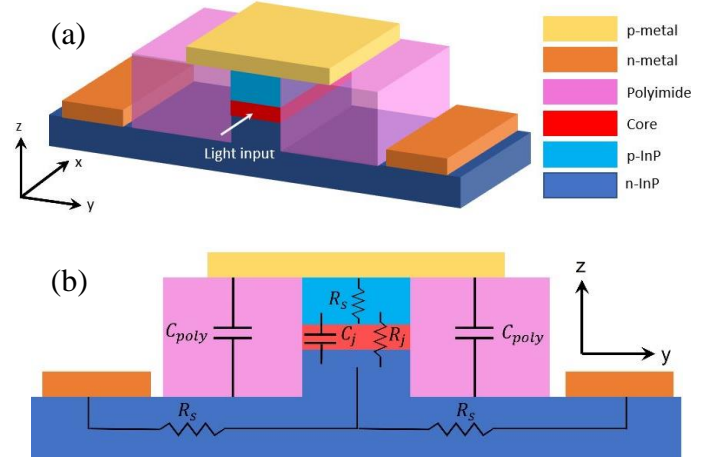

Fig. 2. (a) Bird-view schematic of the detector. Light propagates in *x*-direction. Intrinsic core region, comprising MQW is sandwiched between *p*-InP and *n*-InP. Trenches of waveguide are filled with polyimide. The p-metal is deposited on top of *p*-InP and polyimide. (b) Cross-section view of the detector. Capacitive and resistive components are shown, which are described in the text. Both the schematics are not in scale.

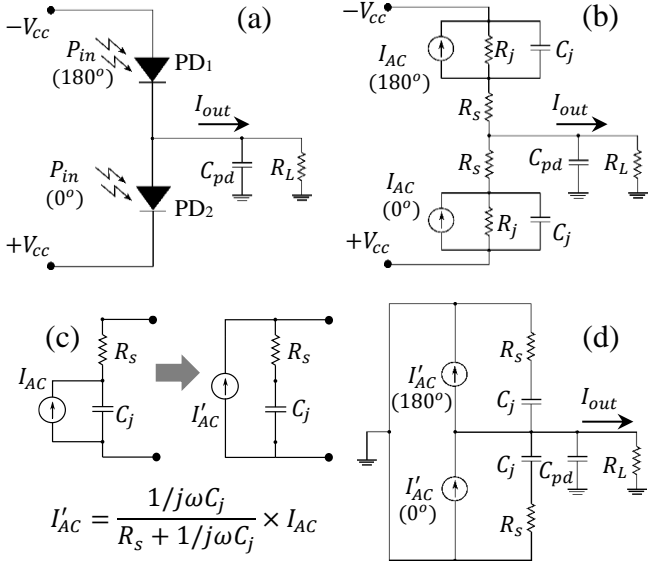

Fig. 3. (a) Schematic of a balanced detector. Light inputs are fed to two photodiodes separately. (b) Equivalent circuit for the balanced detector. (c) Norton equivalent circuit of the active region. Transfer function for the Norton equivalent current source is presented. (d) Equivalent circuit for AC analysis. DC voltage sources are shorted. (a,b,d) are reprinted with permission from [24] © The optical Society.

### III. BANDWIDTH AND RESPONSIVITY CALCULATION

Responsivity of a PD is defined as the amount of current generated with respect to DC optical power, denoted as $R_{DC}$, which can be expressed with the absorption coefficient of QWs in MQW detector. Whereas, bandwidth of PD is defined as the frequency of optical signal where its responsivity drops to 3 dB of $R_{DC}$ [20]. We express the effective frequency-dependent responsivity $R(\omega)$ of BPD with two parts: (i) DC component to know the responsivity and (ii) AC component to compute its bandwidth. This section describes both the components of $R(\omega)$ and corelate them with the physical parameters of BPD.

#### A. DC Responsivity

DC component of responsivity $R_{DC}$ for a waveguide-based PDs depends on different parameters e.g. optical confinement, distance travelled by light, optical absorption at different regions and photon energy. For a known waveguide structure, these parameters can be reliably simulated and/or measured.

Absorbed light in the QWs contribute to the photocurrent. Free-carrier absorption of light in $p$-InP and $n$-InP regions results in heat generation through phonon emission [25]. Light intensity drops exponentially as it travels through an absorbing material [26]. $R_{DC}$ for a BPD of length $L_{PD}$ can be expressed as:

$$R_{DC} = \frac{q}{\hbar\omega} \times e^{-\alpha_n \times \Gamma_n \times L_{PD}} \times e^{-\alpha_p \times \Gamma_p \times L_{PD}} \\ \times (1 - e^{-\Gamma_{QW} \times \alpha_{QW} \times L_{PD}}) \quad (1)$$

Here, length of BPD is denoted by $L_{PD}$. Photon energy of input optical signal is denoted as $\hbar\omega$ and electronic charge is $q$. Optical confinement in the QWs is denoted as $\Gamma_{QW}$. Absorption coefficient of input light for the corresponding wavelength in the QWs is $\alpha_{QW}$. Absorption coefficients for $n$-InP and $p$-InP are denoted as $\alpha_n$ and $\alpha_p$, respectively, which depend on the corresponding doping densities [27]. The optical confinement factors in $n$-InP and $p$-InP regions (excluding the extended depletion regions in reverse bias) are $\Gamma_n$ and $\Gamma_p$, respectively.

#### B. Effective Responsivity

Effective responsivity $R(\omega)$ of a BPD depends on both the optical power and its frequency. It has both the DC (i.e. $R_{DC}$) and AC component. $R(\omega)$ can be expressed as:

$$R(\omega) = \frac{I_{out}}{P_{in}(\omega)} = \frac{I_{out}}{I'_{AC}} \times \frac{I'_{AC}}{I_{AC}} \times \frac{I_{AC}}{P_{in}(\omega)} \quad (2)$$

Three factors in the R.H.S. of equation (2) come from the equivalent circuit of BPD, shown in fig. 3(d). $I_{out}/I'_{AC}$ is the output current gain with respect to Norton equivalent current. Gain ratio of Norton equivalent current to the original current is $I'_{AC}/I_{AC}$ in fig. 3(c). These two factors possess information of bandwidth due to RC loading. Ratio of generated current to the incident optical power is $I_{AC}/P_{in}(\omega)$, that contains information of optical absorption in the QWs and bandwidth limitation due to carrier transit time. By applying node analysis to the equivalent circuit, we solve $I_{out}/I'_{AC}$. Rate equations are solved in the active region to get expression for $I_{AC}/P_{in}(\omega)$. So we get:

$$R(\omega) \propto \frac{\frac{1}{j\omega C_j}}{R_s + \frac{1}{j\omega C_j}} \times \frac{1}{1 + j\omega\tau} \times 2\left(R_s + \frac{1}{j\omega C_j}\right) \times \\ \frac{1}{R_s(1 + j\omega R_L C_{pd}) + 2R_L + \frac{1}{j\omega C_j} \times (1 + j\omega R_L C_{pd})} \quad (3)$$

Elaborated calculations of $R(\omega)$ are described in Appendix A. We do not take DC parameters in the factor $I_{AC}/P_{in}(\omega)$ of equation (2) into account as they will be expressed as $R_{DC}$. For $\omega \to 0$, $R(\omega)$ must lead to $R_{DC}$. We thus normalize $R(\omega)$ for $\omega \to 0$ and express the effective responsivity in equation (4). The DC responsivity $R_{DC}$ is calculated by applying $\omega \to 0$ in equation (4) and bandwidth is calculated as the value of $\omega$ where $20\log_{10}|R(\omega)|$ drops by 3 dB. Effective transit time of

$$R(\omega)$$
$$= R_{DC} \times \frac{\frac{1}{j\omega C_j}}{R_s + \frac{1}{j\omega C_j}} \times \frac{1}{1 + j\omega\tau} \times \left(R_s + \frac{1}{j\omega C_j}\right)$$
$$\times \frac{1}{R_s(1 + j\omega R_L C_{pd}) + 2R_L + \frac{1}{j\omega C_j} \times (1 + j\omega R_L C_{pd})} \quad (4)$$

carriers ($\tau_c$) is included in the term $\tau$. Bandwidth can be limited due to high $\tau_c$ or high RC value. Based on these reasons of limitation, BPD can either be in transit time limited or RC limited region, respectively. Transit time of electrons ($\tau_n$) and holes ($\tau_p$) are calculated as weighted average of optical confinement at $i^{th}$ quantum well and time taken by carriers to travel from that well to the corresponding collecting layer:

$$\tau_n = \frac{\sum_i(\tau_i^n \times \Gamma_i)}{\sum_i \Gamma_i} \quad (5)$$

$$\tau_p = \frac{\sum_i(\tau_i^p \times \Gamma_i)}{\sum_i \Gamma_i} \quad (6)$$

Confinement of optical mode in $i^{th}$ well is $\Gamma_i$. Collection time for electrons and holes generated in the $i^{th}$ well are $\tau_i^n$ and $\tau_i^p$, respectively. Throughout the paper, it is assumed that enough reverse bias is applied to the BPD so that velocity of carriers saturates. Effective transit time ($\tau_c$) can be computed from the effective saturation velocity of carriers in the active region. The saturation velocity ($V_{sat}$) can be expressed as [14]:

$$V_{sat} = \left(0.5 \times \left(\left(\frac{1}{V_{sat}^p}\right)^4 + \left(\frac{1}{V_{sat}^n}\right)^4\right)\right)^{-\frac{1}{4}} \quad (7)$$

Saturation velocity for electrons and holes are $V_{sat}^n$ and $V_{sat}^p$, respectively. Considering effective transit time $\tau_c = d_{act}/V_{sat}$, we can get from equation (5)−(7):

$$\tau_c = \left(0.5 \times (\tau_p^4 + \tau_n^4)\right)^{\frac{1}{4}} \quad (8)$$

## IV. RESULT AND DISCUSSION

This section starts with computing the confinement factor in waveguide. We then present bandwidth of BPDs and discuss bottlenecks to achieve high bandwidth due to various physical parameters. The dimensions we consider to tune are: width of the active region ($d_{act}$), length of the detector ($L_{PD}$), distance of MQWs from $p$-InP (p-offset) and quantum barrier thickness. Optimization of BPDs by our developed PSO-based algorithm is demonstrated for the above-mentioned physical parameters. A design methodology of coherent receiver is also investigated. All simulation parameters are listed in table I.

TABLE I
PARAMETERS FOR THE SIMULATION [14], [28]

| Symbol | Description | Value |
|---|---|---|
| $V_{sat}^p$ | saturation velocity of holes | 6.6×10⁴ m/s |
| $V_{sat}^n$ | saturation velocity of electrons | 2.6×10⁵ m/s |
| $E_{sat}$ | Electric field for velocity saturation | 70 kV/cm |
| $C_{pd}$ | pad capacitance | 12 fF |
| $\epsilon_{act}$ | relative permittivity of active region | 13 |
| $\epsilon_0$ | permittivity of vacuum | 8.854×10⁻¹² F/m |
| $\epsilon_{poly}$ | relative permittivity of polyimide | 2 |
| $n_{core}$ | Refractive index of core | 3.38 |
| $n_{clad}$ | Refractive index of cladding | 3.17 |
| $n_{poly}$ | Refractive index of polyimide | 1.5 |
| $\lambda_0$ | wavelength in vacuum | 1550 nm |
| $\alpha_{QW}$ | absorption coefficient of QWs | 10⁴ cm⁻¹ |
| $\hbar$ | Reduced Plank's constant | 1.055×10⁻³⁴ J-s |
| $n_g$ | Group index | 3.65 |
| $c_o$ | Speed of light in vacuum | 3×10⁸ m/s |
| $\tau_r$ | radiative lifetime | 1 ns |
| $d_{poly}$ | depth of polyimide | 2 $\mu$m |
| $W_{PD}$ | width of detector | 2 $\mu$m |
| $R_s$ | series resistance | 20 Ω |
| $R_L$ | load resistance | 50 Ω |
| $N_A$ | Doping density of p-InP cladding | 10¹⁷ cm⁻³ |
| $N_D$ | Doping density of n-InP cladding | 10¹⁷ cm⁻³ |
| n.i.d | Non intentional n-doping of core | 10¹⁶ cm⁻³ |
| QW | Quantum well thickness | 5 nm |

### A. Optical Confinement in Detectors

Responsivity in waveguide detectors greatly depends on the optical confinement along its physical dimensions. Here, we try to find out how optical confinement varies with width of the BPD ($W_{PD}$). Fig. 4(a−d) show fundamental TE mode in the waveguide of detectors having a width ($W_{PD}$) of 1.5−3.0 $\mu$m using LUMERICAL mode solver. Thickness of the core ($d_{act}$) is kept as 500 nm in all the cases. Insignificant change in the corresponding confinement factors ($\Gamma_{2D}$) shows that the optical field is well confined in *y*-direction for the above range due to large dimension and high contrast in refractive index. We choose this range because the lower limit of $W_{PD}$ is defined by the generic technology. And, very wide waveguide results in lossy higher order modes. But optical confinement depends significantly on $d_{act}$ for the range of 100−600 nm as lower contrast in refractive indices along *z*-direction leads to leakage of light. We calculate confinement factor for *z*-direction of PDs by solving Maxwell equations to find profile of fundamental TE mode [29]. Details of the equations are in Appendix B.

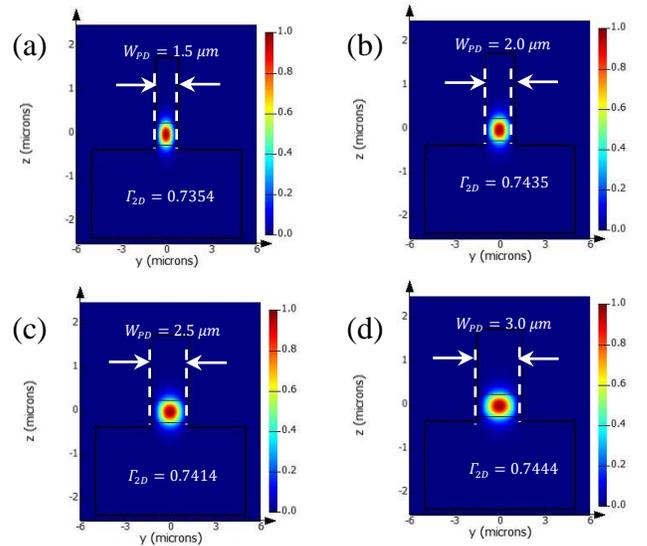

Fig. 4. (a−d) show fundamental TE mode for waveguide PDs with width ($W_{PD}$) in the range 1.5−3.0 $\mu$m. Corresponding confinement factors show insignificant change because of change in $W_{PD}$ as optical confinement is already very high in $y$-direction.

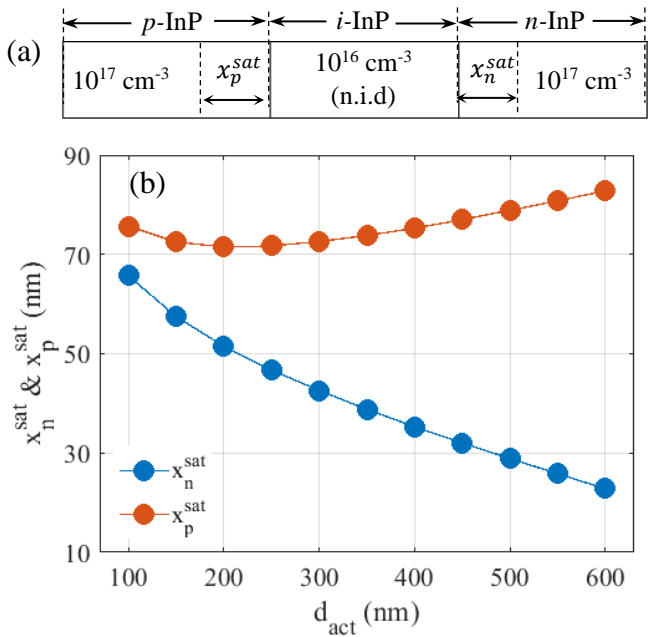

Fig. 5. (a) Schematic of a *p-i-n* diode with relevant doping densities. $x_n^{sat}$ and $x_p^{sat}$ are encroached depletion region in n-InP and p-InP layers, respectively for the reverse bias, required for velocity saturation. (b) $x_n^{sat}$ decreases with increase in $d_{act}$. Whereas, $x_p^{sat}$ shows slight increase.

### B. Calculation of Depletion Width

Calculating depletion width is essential to find out bandwidth of the detectors. Electric field of ~70 kV/cm is required in the active region of InP/InGaAs detectors for velocity saturation of carriers [30]. Required reverse bias to attain the electric field leads to extension of depletion region beyond the intrinsic core region. We calculate the corresponding extension of depletion region into the *n*-InP ($x_n^{sat}$) and *p*-InP ($x_p^{sat}$) layers through depletion approximation [31]. Fig. 5 shows that $x_n^{sat}$ decreases significantly with $d_{act}$. Whereas, $x_p^{sat}$ increases slowly after initial dip. This implies lesser extension of depletion region in *p-i-n* BPDs with thicker cores. We have taken $x_n^{sat}$ and $x_p^{sat}$ into account for calculating junction capacitance and transit time of photogenerated carriers. Encroached depletion regions in *p*-InP and *n*-InP layers are excluded from the calculation of free-carrier absorption, discussed in section III.A.

### C. BPD Optimization

In order to optimize BPD, we need to identify effect of various parameters on bandwidth and responsivity. Equation (4) is solved to calculate bandwidth and responsivity. Fig. 6 shows bandwidth for different p-offset and n-offset at different lengths ($L_{PD}$) of BPD. Thickness of the quantum barrier is considered as 10 nm in both the sub-figures. Core thickness ($d_{act}$) is varied by changing only the p-offset and n-offset, respectively. It can be observed from fig. 6(a) that depending on p-offset and $L_{PD}$, BPD falls into two regions: transit time limited regime because of higher values of p-offset and RC limited regime due to higher capacitance from thinner p-offset. Also, longer BPD leads to lower bandwidth. This comes from equation (4), as explained in section III. Fig. 6(b) shows that higher values of n-offset don't push BPD into transit time limited regime unlike for the case of higher p-offset in fig. 6(a). Much faster electrons essentially keep BPD in the RC limited region. MQW stack is sufficiently close to *p*-InP for p-offset of 50 nm. So, the slower holes cannot put the BPD into transit time limited regime. Evolution of BPD bandwidths with barrier thickness for different p-offset values are shown in fig. 7. $L_{PD}$ is calculated at each datapoint to satisfy the responsivity of 1.0 A/W. For $d_{act} = 400$ nm in fig. 7(a), p-offset of 50 nm places the MQW layerstack too close to the *p*-InP layer. In this case, confinement of optical mode in the quantum wells is low for too thin or too thick quantum barrier, as optical mode is mostly confined in the middle of the core (fig. 4). Thus, longer BPD is required to achieve the desired responsivity, pushing the BPD into RC limited regime. However, p-offset of 150 and 200 nm means that the MQW stack is too far from $p$-InP, resulting in lesser bandwidth due to high transit time of slow holes. So, highest bandwidth can be obtained for an optimum p-offset. Similar explanation can be deduced for fig. 7(b) in case of $d_{act}=600$ nm and it is preferred that MQW stack stays close to $p$-InP to obtain high bandwidth irrespective of the core thickness ($d_{act}$). We can conclude from fig. 7 that barrier layer between QWs should be of minimum thickness for the optical confinement factor to be maximum. However, it should be thick enough to prevent coupling between quantum wells [32]. So, we can tune remaining three parameters (i.e. $d_{act}$, $L_{PD}$, and p-offset) to optimize a BPD, which is done through a PSO-based algorithm. Details of the algorithm are provided in the Appendix C and parameters are chosen accordingly [33].

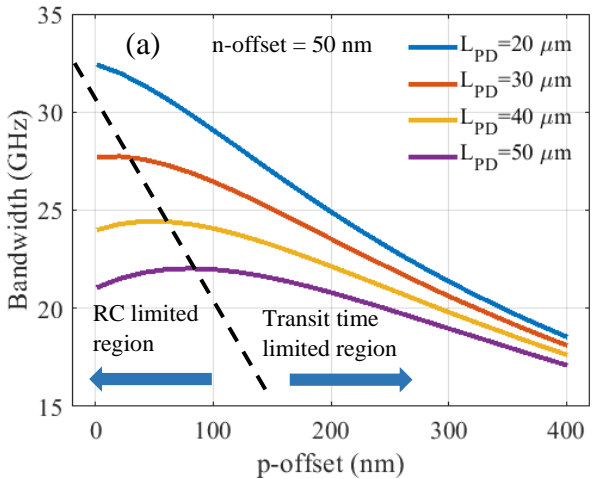 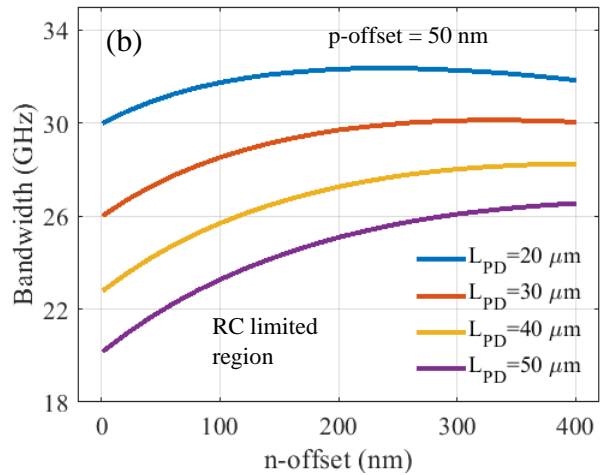

Fig. 6. (a) and (b) show bandwidth of BPD vs. p-offset and n-offset, respectively for different lengths of BPD. Higher p-offset is more likely to push the BPD into transit time limited regime as opposed to n-offset as holes are slower. Shorter BPD has higher bandwidth due to lesser RC value.

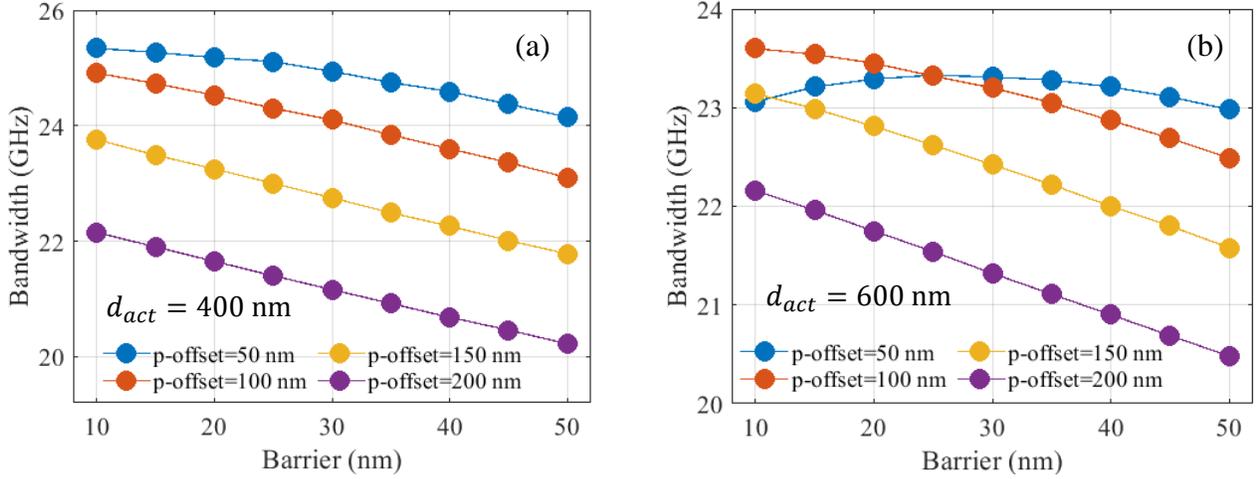

Fig. 7. Evolution of bandwidth of BPD with barrier thickness for different p-offset values. Length of the BPD is chosen to make responsivity to be 1.0 A/W at each data point. $d_{act}$ is taken as 400 and 600 nm in (a) and (b), respectively. Highest bandwidth is achieved for the lowest barrier thickness at an optimum p-offset irrespective of $d_{act}$.

### D. Optimization by PSO Algorithm

In order to design BPD for the highest bandwidth at a given responsivity of 1.0 A/W, we choose 40 particles with random initial positions of three dimensions with reference to the three variables i.e. $d_{act}$, $L_{PD}$ and p-offset within the user-defined solution space. The PSO algorithm calculates bandwidth and responsivity using equation (4) at each iteration. Every particle moves according to their position and velocity to check for the highest bandwidth if it has achieved the desired responsivity. Flowchart of the algorithm is presented in fig. 8. Standard deviations of three variables decrease with iterations indicating convergence of the particles, as shown in Fig. 9. The initial overshoot comes from high particle velocity in order to make the search process effective. The algorithm is able to find optimum BPD design depending on the trade-off between RC limited and transit time limited regions. Multiple combinations of design parameters are possible to achieve almost the same bandwidth at the same responsivity. Thus, the algorithm has a potential to provide different optimum designs of BPD, as presented in table II. Some of these designs maybe easier to grow or fabricate. The values of bandwidth and responsivity don't deviate much from their mean values for 10 samples, having standard deviations of ~0.05 GHz and ~0.006 A/W, respectively. Compared to single $p$-$i$-$n$ PD, BPD suffers from reduced bandwidth due to two junction capacitances in parallel [34]. The last column of table II (i.e. $\Delta$) indicates the distance from *p*-InP layer to the midpoint of MQW stack. Total thickess of MQW stack is 50 nm. So, in all the cases, MQW stack stays closer to the *p*-InP layer to get the highest bandwidth so that it is not limited by slower holes. Section IV.F dsicusses which of these BPD designs to be picked for designing coherent receiver.

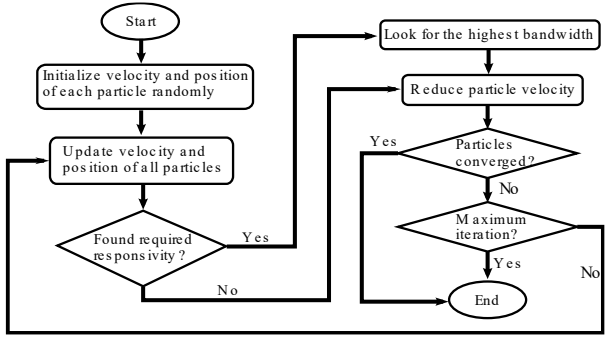

Fig. 8. Flowchart of the PSO algorithm to optimize BPDs. Particles are initialized with random position and velocity. Then each iteration updates their position and velocity to search for the highest bandwidth if the responsivity is already met. This process goes on until all the particles converge or maximum iteration reached.

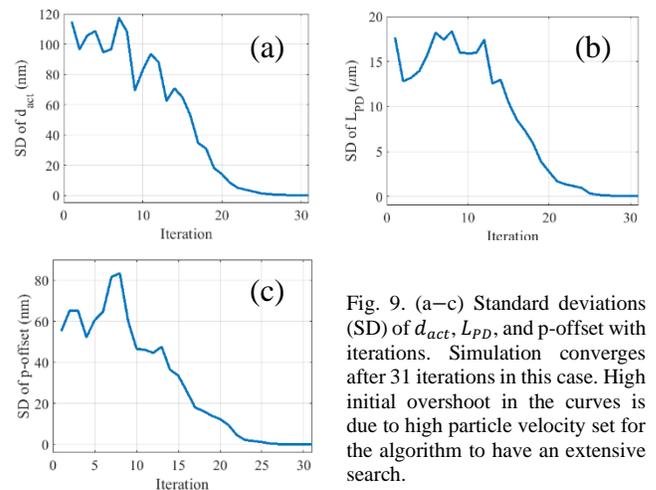

Fig. 9. (a–c) Standard deviations (SD) of $d_{act}$, $L_{PD}$, and p-offset with iterations. Simulation converges after 31 iterations in this case. High initial overshoot in the curves is due to high particle velocity set for the algorithm to have an extensive search.

TABLE II
COMBINATION OF PARAMETERS BY PSO ALGORITHM

| BW (GHz) | R (A/W) | $d_{act}$ (nm) | $L_{PD}$ ($\mu$m) | p-offset (nm) | $\Delta$ (nm) |
|---|---|---|---|---|---|
| 25.4 | 1.0014 | 350.55 | 51.96 | 46.35 | 71.35 |
| 25.4 | 1.0008 | 369.41 | 52.67 | 47.14 | 72.14 |
| 25.4 | 1.0000 | 371.99 | 50.56 | 63.11 | 88.11 |
| 25.4 | 1.0012 | 379.70 | 52.93 | 49.08 | 74.08 |
| 25.4 | 1.0006 | 336.59 | 51.02 | 47.36 | 72.36 |
| 25.3 | 1.0005 | 384.11 | 51.17 | 63.63 | 88.63 |
| 25.4 | 1.0003 | 381.89 | 52.37 | 53.39 | 78.39 |
| 25.3 | 1.0020 | 377.88 | 53.69 | 46.19 | 71.19 |
| 25.4 | 1.0002 | 352.87 | 51.35 | 51.05 | 76.05 |
| 25.3 | 1.0007 | 364.18 | 49.50 | 69.14 | 94.14 |

### E. Bandwidth vs. Various Parameters

Optimum performance of BPD has also been investigated for a number of parameters using the developed algorithm. In this sub-section, we only present maximum possible bandwidth for a range of responsivities $(R)$, width of the detector $(W_{PD})$ and absorption coefficient of QWs $(\alpha_{QW})$ at the wavelength of 1550 nm. Optimum BPD designs (similar to table II) for each responsivity are not shown to keep the text simple. Fig. 10 shows that the highest bandwidth of BPD decreases with higher responsivity. This can be attributed to two requirements of higher responsivity: (i) longer detector, that increases RC value and (ii) thicker active region, leading to higher transit time. Fig. 11 shows that the highest bandwidth decreases with the width of BPD. This is because of higher contribution of junction capacitance $(C_j)$ to effective capacitance $(C_{eff})$ as permittivity of the core region is higher than that of polyimide. So, BPDs with narrow waveguides give high bandwidths due to lesser capacitance. But, it is subject to technological limitations and high optical scattering due to sidewall roughness [35]. QW design can be engineered to get stronger absorption. So, we run the algorithm to determine how absorption coefficient of QWs $(\alpha_{QW})$ affects the performance of BPD, as shown in fig. 12. Increase of the highest bandwidth with $\alpha_{QW}$ is due to the fact that higher $\alpha_{QW}$ allows shorter detector and thinner absorbing layer for the same responsivity. It makes the BPD less impacted by RC loading and transit time. In both the cases of figure 11 and 12, changes are more noticeable for lesser bandwidth.

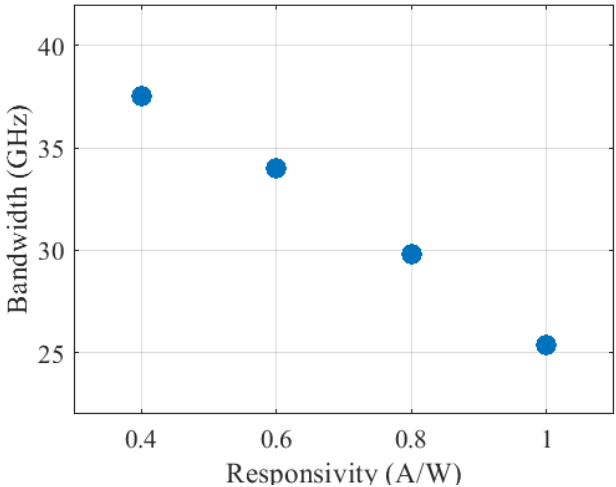

Fig. 10. Highest bandwidth of BPD vs. responsivity values.

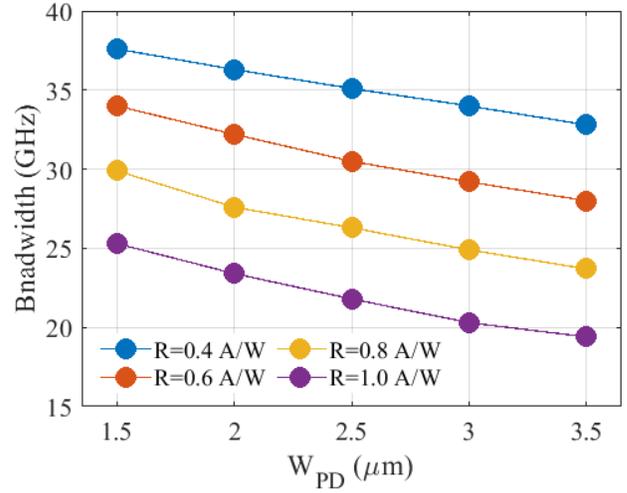

Fig. 11. Highest bandwidth of BPD vs. BPD width $(W_{PD})$ for different values of responsivity (R).

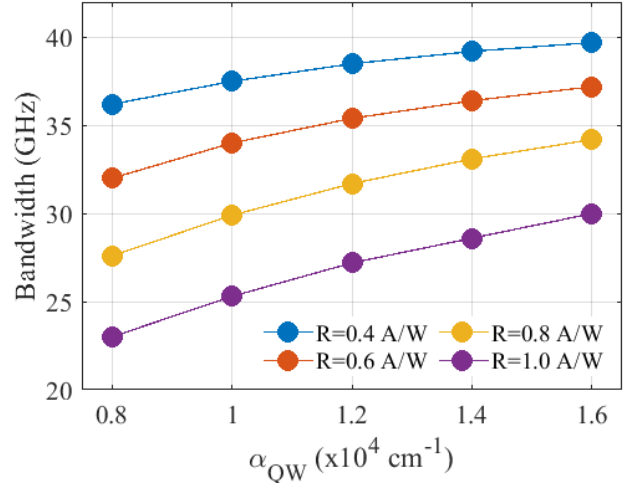

Fig. 12. Highest bandwidth of BPD vs. absorption coefficient of QWs $(\alpha_{QW})$ for different values of responsivity (R).

### F. Coherent Receiver Design: A Case Study

This work aims to optimize BPD for coherent receiver on generic InP platform, as mentioned in section I. Fig. 13 shows a simple schematic of photonic integrated circuit (PIC) part of a coherent balanced receiver [36]–[38]. Optical power of laser and input are denoted as L and S, respectively. We focus on a heterodyne coherent receiver as it is less affected by phase noise and is suitable for complex modulation schemes such as QAM and DPSK [39]. Control signals of the laser and the phase shifter come from a phase-locked loop (PLL) and are not shown here. Difference current of two detectors in a BPD of coherent receiver is further processed by a transimpedance amplifier (TIA), which demands a certain amount of current from BPD for successful detection. This depends on the power level of input signal, laser output, responsivity of the BPD and losses in the circuit. Minimum required current output current of BPD is considered to be 160 $\mu A_{pp}$ (difference of current between positive and negative maxima), which aligns with linear TIAs, investigated by Awny et al. [40]. Input signal and laser output are mixed in a 90° hybrid coupler and four output signals are then fed to the two sets of BPDs. Phases of two optical signals, going to a pair of detectors in a BPD are 180° apart [36]–[38]. Resultant photocurrent of each PD is directly related to the square of effective incident electric field [39]. The required responsivity of BPDs to meet minimum current requirement of TIA for a range of input power are shown in fig. 14 for different laser power levels. Appendix D provides methodology for this calculation. For a higher laser and/or signal power, required responsivity of BPD goes lower. This can lead to a higher bandwidth, as demonstrated in figure 10. But, there is a limit in laser power to prevent thermal runaway, detector nonlinearity, saturation of output current, violation of power handling limit by the PDs. Input optical power is considered to be the lowest value expected during the operation of receiver. So, we need to optimize design parameters of BPD for a certain combination of laser and input power while considering all lossy elements on receiver circuit.

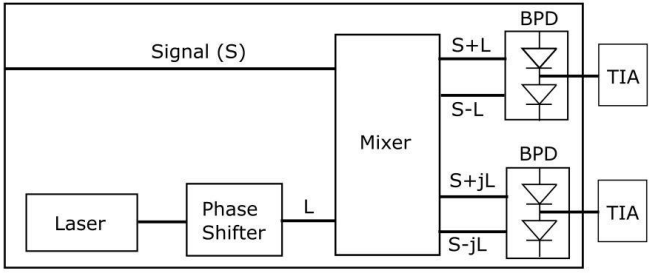

Fig. 13. Simple schematic of a coherent balanced receiver.

The developed PSO algorithm is able to produce several optimized BPD designs. To realize the coherent receiver, we need to find out which of these designs can be monolithically integrated with laser, that works as the local oscillator. Let's consider a scenario, where we know threshold current for lasing $I_{th}$, laser power at a corresponding driving current and position of MQW stack in the waveguide core on generic platform. The power budget estimation of coherent receiver tells us how much of laser power can be reduced and the corresponding required BPD responsivity, as presented in section II.A. The required responsivity of BPD must be within the theoretical limit at the wavelength of input optical signal (e.g. 1.25 A/W for 1550 nm). Combinations of different p-offset and $d_{act}$ for BPD shall be determined through the algorithm for the highest bandwidth at that required responsivity. Considering the same layerstack for laser, threshold current for lasing ($I_{th}$) can then be calculated for a specific set of p-offset and $d_{act}$. Here, we calculate $I_{th}$ and optical confinement in QWs for a wide range of p-offset at different $d_{act}$. Simulation parameters are taken from previous works on generic platform [41], [42]. Table II shows that MQW should be closer to *p*-InP for optimum BPD. But it degrades confinement in the QWs, which increases the threshold current for lasing, as shown in fig. 15. It can be explained with the

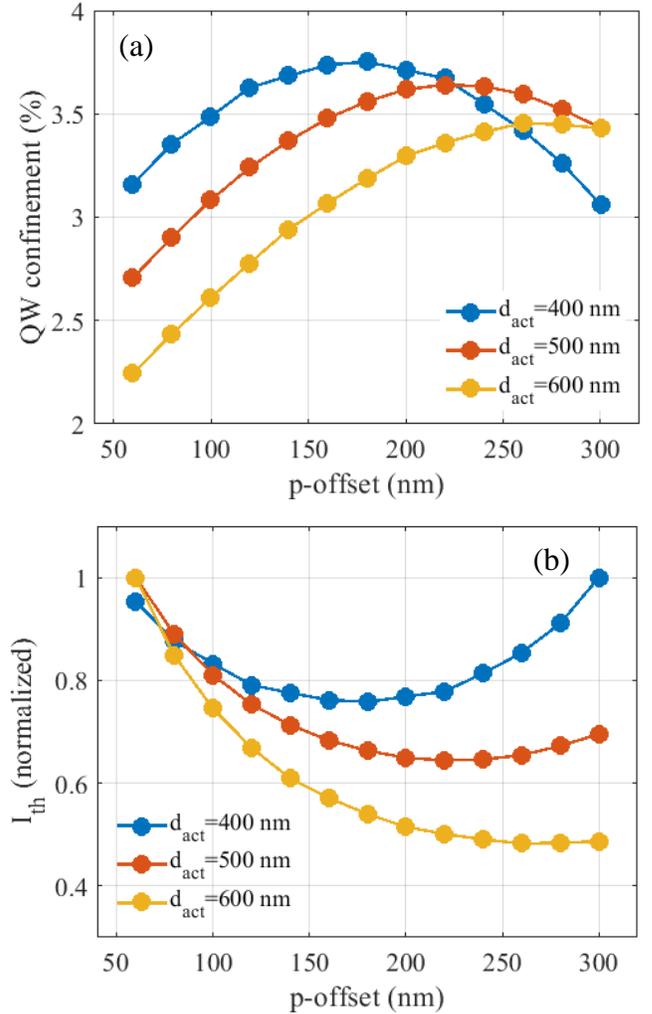

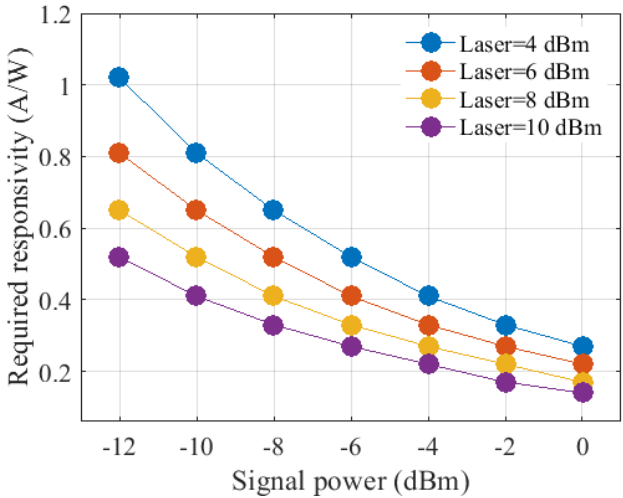

Fig. 14. Required responsivity of balanced detector in the coherent receiver. It increases for low signal power and/or low laser power.

Fig. 15. Optical confinement in QWs is the highest and threshold current of laser ($I_{th}$) is at its minimum if the MQW stack is in the middle of active region, shown in (a) and (b), respectively.

following condition for lasing [13]:

$$\Gamma_{QW} g(I) - \alpha_L = 0 \quad (9)$$

Optical confinement factor of QWs is $\Gamma_{QW}$. Material gain is $g(I)$ which is dependent on driving current $I$ and gets clamped at the onset of lasing. It is considered that the injected current is not too high and laser operates in linear region. Total cavity loss is denoted as $\alpha_L$. Higher current injection is needed to fulfil the equation (9) for lower $\Gamma_{QW}$, which increases the threshold current. Since laser power increases linearly with $I - I_{th}$, it with also decrease with higher $I_{th}$ when MQW stack deviates from the middle of core region. Appendix E provides details of this simulation. Together with fig. 15(b) and measured laser power, the developed PSO-based algorithm will be able to guide us which of the optimized BPD designs should be implemented in the generic platform while making sure of the required laser power for coherent receiver.

## V. CONCLUSION

To conclude, we have developed a design methodology to optimize MQW BPDs on generic InP platform through a PSO-based algorithm using its equivalent electrical circuit. We have designed BPD for the highest bandwidth possible at a given responsivity while making sure of generic integration. Impact of different physical parameters on BPD's performance are thoroughly investigated and strategies to circumvent major bottlenecks have been identified. Quantum barriers must be as thin as possible without leading to coupling of QWs. Values of corresponding design parameters i.e. length and width of BPD, and core thickness are optimized via the PSO algorithm. The highest bandwidth increases with lower responsivity, thinner detectors and higher absorption coefficient of QWs. The MQW stack should be closer to p-InP layer to achieve high bandwidth by overcoming the bottleneck of slower holes. But, using the same layerstack of BPD for laser with MQW stack closer to p-InP leads to higher threshold current for lasing. So, a trade-off has to be made, which is presented in a case study of designing a coherent receiver on generic InP platform through successful integration of optimized BPD with laser.

## APPENDIX

### A. Deriving Components of R(ω)

Fig. 16 shows the equivalent circuit of BPD for AC analysis along with current components through different branches. Analyzing KCL and KVL at node B and C give us:

$$I'_{out} = 2I'_{AC} - 2I_2 \quad (10)$$

$$I'_{out} \times \left(\frac{R_L}{1 + j\omega R_L C_{pd}}\right) = I_2 \times \left(R_s + \frac{1}{j\omega C_j}\right) \quad (11)$$

$$I_{out} = \frac{1/j\omega C_{pd}}{R_L + 1/j\omega C_{pd}} \times I'_{out} \quad (12)$$

Solving equations (10)−(12) gives:

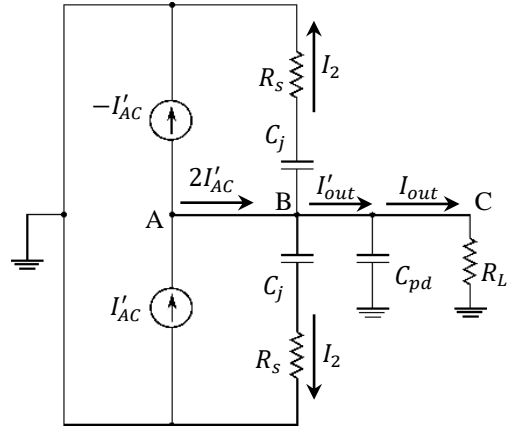

Fig. 16. Equivalent circuit of BPD for AC analysis. Relations between current components through different branches are in the text.

$$\frac{I_{out}}{I'_{AC}} = \frac{2\left(R_s + \frac{1}{j\omega C_j}\right)}{R_s(1 + j\omega R_L C_{pd}) + 2R_L + \frac{1 + j\omega R_L C_{pd}}{j\omega C_j}} \quad (13)$$

In order to find out the term $I_{AC}/P_{in}$ of equation (1), we need to consider dynamic behavior of the PD. Carrier dynamics in the PD can be expressed with the following rate equation:

$$\frac{dn}{dt} = Gn_0 P - \frac{n}{\tau} \quad (14)$$

The 1st term on R.H.S. is generation rate of carriers, which depends on intrinsic carrier density ($n_0$), photon density ($P$) and generation coefficient $G$. The 2nd term is the carrier removal rate, where $\tau$ is expressed as,

$$\frac{1}{\tau} = \frac{1}{\tau_c} + \frac{1}{\tau_r} \quad (15)$$

Here, $\tau_c$ is the carrier transit time after generation through applied electric field and $\tau_r$ is the recombination lifetime. We express photocurrent in terms of carrier sweep-out rate as,

$$I_{AC} = \frac{qnV}{\tau_c} \quad (16)$$

Here, $q$ is the electronic charge and $V$ is the volume of active region. From the equations (14)−(16),

$$\tau \times \frac{dI_{AC}}{dt} + I_{AC} = \frac{qV\tau Gn_0 P}{\tau_c} \quad (17)$$

Considering $I_{AC} = I_0 \times \exp(j\omega t)$, equation (17) can be expressed as:

$$I_{AC} = \frac{qV\tau Gn_0 P}{(1 + j\omega\tau)\tau_c} \quad (18)$$

Considering that photon density $P$ is linearly dependent on input optical power $P_{in}$, it can be written from equation (18):

$$\frac{I_{AC}}{P_{in}(\omega)} \propto \frac{1}{1+j\omega\tau} \quad (19)$$

### B. Calculating Optical Confinement in Waveguide

Electric field profile for TE mode along z-direction (see fig. 1) is calculated from the formula:

$$\frac{d^2 E_z}{dy^2} + [k_0^2 n^2(z) - \beta^2] E_z = 0 \quad (20)$$

Here, $\beta$ is propagation constant, $n(z)$ is the refractive index profile along z-direction and $k_0$ is the wavenumber. Optical confinement in the QWs is calculated as:

$$\Gamma_z = \frac{\int_{QW} |E_z|^2 dz}{\int_{-\infty}^{\infty} |E_z|^2 dz} \quad (21)$$

### C. Implementing Particle Swarm Optimization

Velocity and position of each particle are updated as per the following formulae:

$$V_{d,L}(i+1) = w \times V_{d,L}(i) + C_1 \left( P_{best}^{d,L} - P_{d,L}(i) \right) + C_2 \left( G_{best}^{d,L} - P_{d,L}(i) \right) \quad (22)$$

$$P_{d,L}(i+1) = P_{d,L}(i) + V_{d,L}(i) \quad (23)$$

The first term on R.H.S of equation (22) depends on already existent velocity of the particle. Second term is the velocity component driven by the particle's best position so far. And the third term is determined by the best find so far by the whole swarm. The parameter w is inertia coefficient, $C_1$ is cognitive coefficient and $C_2$ is social coefficient. $P_{best}$ is the personal best position of the corresponding particle and $G_{best}$ is the global best position found by the swarm so far. The values of $w$ and $C_1$ are reduced after every iteration by a damping factor $(f_{damp} < 1)$ as $w(\text{or } C_1) = w(\text{or } C_1) \times f_{damp}$. The value of $C_2$ remains undamped, otherwise we observe particles to freeze after a few iterations. Damping only $w$ and $C_1$ make sure that the whole swarm converges towards the global best.

### D. Calculations for Heterodyne Coherent Receiver

The optical field associated with input signal and laser output can be expressed as:

$$E_s = A_s \exp[-i(\omega_0 t + \phi_s)] \quad (24)$$

$$E_{LO} = A_{LO} \exp[-i(\omega_{LO} t + \phi_{LO})] \quad (25)$$

$A_s$, $\omega_0$ and $\phi_s$ represent amplitude, frequency and phase of input signal, respectively. Whereas, the same parameters for laser are $A_{LO}$, $\omega_{LO}$ and $\phi_{LO}$, respectively. The optical power on PDs is $P = K|E_s + E_{LO}|^2$, where $K = 0.5 \times \sqrt{\epsilon_0 \epsilon_r / \mu_0} \times A_{WG}$. Area of the cross-section of PD waveguide is $A_{WG}$. So, $P$ can be expressed as:

$$P(t) = P_s + P_{LO} + 2\sqrt{P_s P_{LO}} \cos(\omega_{IF} t + \phi_s - \phi_{LO}) \quad (26)$$

$P_s = KA_s^2$, $P_{LO} = KA_{LO}^2$ and $\omega_{IF} = \omega_0 - \omega_{LO}$. We consider $\omega_{IF} = 1$ GHz for heterodyne receiver. Phase difference i.e. $|\phi_s - \phi_{LO}|$ will be maintained through the phase shifter and considered to be $\pi/2$ in our case. Current output of a PD is $RP(t)$, where $R$ is responsivity. Current of BPD is defined as the difference currents of two PDs with a 180º phase difference. Propagation losses for input signal and laser for our platform are considered for calculations.

### E. Calculating Threshold Current for Lasing

Equations (27)−(35) have been solved to calculate threshold current of laser $I_{th}$ [13], [41]. Equations (31)−(35) have been solved self-consistently to obtain values for $I_{th}$ and generated photon density $N_p$. Descriptions of parameters in the equations are provided in table III. Effective optical confinement in the SOA is $\Gamma$. Mirror loss and average internal losses are calculated in the equations (28) and (29), respectively. Group velocity is $v_g$. Effective cavity gain is $g'_{0N}$ which reduces due to very high $N_p$, defined by a gain saturation factor $g_c$. Threshold gain and the corresponding threshold carrier density are $g_{th}$ and $N_{th}$, respectively. $V$ is the volume of active medium i.e. QWs. The threshold current for lasing $I_{th}$ doesn't show change even for a driving current of 200 mA. This is because $1 + g_c N_p \cong 1$ and the denominator in equation (31) remains almost unchanged.

$$\Gamma = \Gamma_{QW} \times \frac{L_{SOA}}{L_{SOA} + L_{passive}} \quad (27)$$

$$\alpha_m = \frac{1}{L_{SOA} + L_{passive}} \log\left(\frac{1}{\sqrt{R_1 R_2}}\right) \quad (28)$$

$$\alpha_i = \frac{\alpha_{SOA} L_{SOA} + \alpha_{passive} L_{passive} + \alpha_{excess}}{L_{SOA} + L_{passive}} \quad (29)$$

$$v_g = c_0/n_g \quad (30)$$

$$g'_{0N} = \frac{g_0}{1 + g_c N_p} \quad (31)$$

$$g_{th} = (\alpha_i + \alpha_m)/\Gamma \quad (32)$$

$$N_{th} = N_{tr} \times \exp\left(\frac{g_{th}}{g'_{0N}}\right) \quad (33)$$

$$I_{th} = \frac{qV}{\eta} \times (BN_{th}^2 + CN_{th}^3) \quad (34)$$

$$N_p = \frac{\eta(I - I_{th})}{q v_g g_{th} V} \quad (35)$$

TABLE III
PARAMETERS FOR THE SIMULATION [41], [42]

| Symbol | Description | Value |
|---|---|---|
| $L_{SOA}$ | Length of SOA | 500 $\mu m$ |
| $L_{passive}$ | Length of passive section | 1500 $\mu m$ |
| $R_1$ | Reflectivity rear mirror | 0.92 |
| $R_2$ | Reflectivity of front mirror | 0.85 |
| $\alpha_{SOA}$ | Internal loss of SOA | 25 cm$^{-1}$ |
| $\alpha_{passive}$ | Loss of passive section | 3 dB/cm |
| $\alpha_{excess}$ | Excess loss in the cavity | 3 dB |
| $g_0$ | Material gain of SOA | 2200 cm$^{-1}$ |
| $N_{tr}$ | Transparent carrier density | 5x10$^{17}$ cm$^{-3}$ |
| $B$ | Radiative recombination coefficient | 2.6x10$^{-10}$ cm$^3$s$^{-1}$ |
| $C$ | Auger recombination coefficient | 1.4x10$^{-28}$ cm$^6$s$^{-1}$ |
| $g_c$ | Gain saturation factor | 5x10$^{-17}$ cm$^3$ |
| $\eta$ | Carrier injection efficiency | 0.80 |

REFERENCES

[1] L. T. Nichols, K. J. Williams, and R. D. Esman, "Optimizing the ultrawide-band photonic link," IEEE Transaction on Microwave Theory and Techniques, vol. 45, pp. 1384–1389, Aug. 1997.
[2] M. S. Islam, T. Chau, S. Mathai, T. Itoh, M. C. Wu, D. L. Sivco, A. Y. Cho, "Distributed balanced photodetectors for broad-band noise suppression," IEEE transactions on microwave theory and techniques, vol. 47, no. 7, pp. 1282–1288, Jul 1999.
[3] C. B. Albert, C. Huang, E. H. Chan, "Intensity noise suppression using dual-polarization dual-parallel modulator and balanced detector," IEEE Photonics Journal, vol. 10, no. 2, pp. 1–8, Nov 2017.
[4] X. Wang, N. Wada, T. Miyazaki, K. I. Kitayama, "Coherent OCDMA system using DPSK data format with balanced detection," IEEE Photonics Technology Letters, vol. 18, no. 7, pp. 826–828, Mar. 2006.
[5] A. H. Gnauck, P. J. Winzer, "Optical phase-shift-keyed transmission," Journal of lightwave technology, vol. 23, no. 1, pp. 115–130, Jan 2005.
[6] Y. Painchaud, M. Poulin, M. Morin, M. Têtu, "Performance of balanced detection in a coherent receiver," Optics express, vol. 17, no. 5, pp. 3659–3672, Mar. 2009.
[7] D. Maes, S. Lemey, G. Roelkens, M. Zaknoune, V. Avramovic, E. Okada, P. Szriftgiser, E. Peytavit, G. Ducournau, B. Kuyken, "High-speed uni-traveling-carrier photodiodes on silicon nitride," APL Photonics, vol. 8, no. 1, Jan. 2023.
[8] Y. Gao, Z. Zhong, S. Feng, Y. Geng, H. Liang, A. W. Poon, K. M. Lau, "High-speed normal-incidence pin InGaAs photodetectors grown on silicon substrates by MOCVD," IEEE Photonics Technology Letters, vol. 24, no. 4, pp. 237–239, Nov. 2011.
[9] H. G. Bach, A. Beling, G. G. Mekonnen, R. Kunkel, D. Schmidt, W. Ebert, A. Seeger, M. Stollberg, W. Schlaak, "InP-based waveguide-integrated photodetector with 100-GHz bandwidth," IEEE Journal of Selected Topics in Quantum Electronics, vol. 10, no. 4, pp. 668–672, Oct. 2004.
[10] M. Smit, X. Leijtens, E. Ambrosius, E. Bente, J. Van der Tol, B. Smalbrugge, T. De Vries, E. J. Geluk, J. Bolk, R. Van Veldhoven, L. Augustin, "An introduction to InP-based generic integration technology," Semiconductor Science and Technology, vol. 29, no. 8, pp. 083001, Jun. 2014.
[11] L. Xu, M. Nikoufard, X. J. Leijtens, T. de Vries, E. Smalbrugge, R. Notzel, Y. S. Oei, M. K. Smit, "High-performance InP-based photodetector in an amplifier layer stack on semi-insulating substrate," IEEE Photonics Technology Letters, vol. 20, no. 23, pp. 1941-1943, Nov. 2008.
[12] L. M. Augustin, R. Santos, E. den Haan, S. Kleijn, P. J. Thijs, S. Latkowski, D. Zhao, W. Yao, J. Bolk, H. Ambrosius and S. Mingaleev, "InP-based generic foundry platform for photonic integrated circuits," IEEE journal of selected topics in quantum electronics, vol. 24, no. 1, pp. 1-10, Jun. 2017.
[13] L. A. Coldren, S. W. Corzine, M. L. Mashanovitch, "Diode lasers and photonic integrated circuits," John Wiley & Sons; Mar. 2012.
[14] G. Ghione, "Semiconductor devices for high-speed optoelectronics," Cambridge: Cambridge University Press, Oct. 2009, ch. 4.
[15] M. S. Islam, T. Jung, T. Itoh, M. C. Wu, A. Nespola, D. L. Sivco, A. Y. Cho, "High power and highly linear monolithically integrated distributed balanced photodetectors," Journal of lightwave technology, vol. 20, no. 2, pp. 285, Feb. 2002.
[16] J. W. Shi, Y. S. Wu, C. Y. Wu, P. H. Chiu, C. C. Hong, "High-speed, high-responsivity, and high-power performance of near-ballistic uni-traveling-carrier photodiode at 1.55-μm wavelength," IEEE photonics technology letters, vol. 17, no. 9, pp. 1929–1931, Aug. 2005.
[17] D. H. Jun, J. H. Jang, I. Adesida, J. I. Song, "Improved efficiency-bandwidth product of modified uni-traveling carrier photodiode structures using an undoped photo-absorption layer," Japanese journal of applied physics, vol. 45, no. 4S, pp. 3475, Apr. 2006.
[18] Y. Liu, K. Jiang, Z. Jiang, B. Zhang, D. Luo, Y. Liu, L. Qu, W. Liu, L. Wang, "High responsivity evanescently coupled waveguide photodiode using spot-size converter and distributed Bragg reflector at 1.55 μm wavelength," Infrared Physics & Technology, vol. 130, pp. 104619, May 2023.
[19] P. Runge, G. A. Zhou, T. Beckerwerth, F. Ganzer, S. Keyvaninia, S. Seifert, W. Ebert, S. Mutschall, A. Seeger, M. Schell, "Waveguide integrated balanced photodetectors for coherent receivers," IEEE Journal of Selected Topics in Quantum Electronics, vol. 24, no. 2, pp. 1–7, Jul. 2017.
[20] A. Beling, J. C. Campbell, "InP-based high-speed photodetectors," Journal of lightwave technology, vol. 27, no. 3, pp. 343–355, Feb. 2009.
[21] P. S. R. Murthy, "Power systems analysis," Butterworth-Heinemann, Jun. 2017, ISBN: 9780081011119, ch. 9.
[22] R. Eberhart, J. Kennedy, "Particle swarm optimization," InProceedings of the IEEE international conference on neural networks, vol. 4, pp. 1942–1948, Nov. 1995.
[23] K. Y. Lee, J. B. Park, "Application of particle swarm optimization to economic dispatch problem: advantages and disadvantages", In2006 IEEE PES power systems conference and exposition, pp. 188–192, Oct 2006.
[24] D. Nag, W. Yao, J. J. Van Der Tol, K. A. Williams, "Investigating RC and Transit Time Limited Bandwidth of Integrated Balanced Detectors through an Equivalent Circuit Model," InOptical Devices and Materials for Solar Energy and Solid-state Lighting, Optica Publishing Group, pp. JTu2A-40, Jul. 2022.
[25] C. Kittel, "Introduction to solid state physics," John Wiley & sons, inc; 2005.
[26] J. Piprek, "Semiconductor optoelectronic devices: introduction to physics and simulation," Elsevier; Oct. 2013.
[27] U. Khalique, "Polarization based integration scheme (POLIS)," (Doctoral dissertation, Ph. D. thesis, Dept. Electr. Eng., Tech. Univ. Eindhoven, Eindhoven, The Netherlands), 2008, ch. 2.
[28] L. Xu, "Monolithic integrated reflective transceiver in indium phosphide," (Doctoral dissertation, Ph. D. thesis, Dept. Electr. Eng., Tech. Univ. Eindhoven, Eindhoven, The Netherlands), 2009, ch. 5.
[29] M. N. Sadiku, S. Nelatury, "Elements of electromagnetics," New York: Oxford university press; 2001, ch. 10.
[30] S. Adachi, "Properties of group-IV, III-V and II-VI semiconductors," John Wiley and Sons, Chichester, England, 2005, ISBN 0-470-09032-4.
[31] R. F. Pierret, "Semiconductor device fundamentals," Pearson Education India; 1996, ch. 5.
[32] B. R. Nag, "Physics of quantum well devices," Springer Science & Business Media, Nov. 2001.
[33] W. Zhang, H. Li, Q. Zhao, H. Wang, "Guidelines for parameter selection in particle swarm optimization according to control theory," InFifth IEEE International Conference on Natural Computation, vol. 3, pp. 520–524, Aug. 2009.
[34] A. Beling, H. G. Bach, D. Schmidt, G. G. Mekonnen, M. Rohde, L. Molle, H. Ehlers, A. Umbach, "High-speed balanced photodetector module with 20dB broadband common-mode rejection ratio," InOptical Fiber Communication Conference, Optica Publishing Group, Mar. 2003.
[35] Y. Wang, M. Kong, Y. Xu, Z. Zhou, "Analysis of scattering loss due to sidewall roughness in slot waveguides by variation of mode effective index," Journal of Optics, vol. 20, no. 2, pp. 025801, Jan. 2018.
[36] Y. Painchaud, M. Poulin, M. Morin, M. Têtu, "Performance of balanced detection in a coherent receiver," Optics express, vol. 17, no. 5, pp. 3659–3672, Mar. 2009.
[37] I. Kaminow, T. Li, A. E. Willner, "Optical fiber telecommunications VB: systems and networks," Elsevier; Jul. 2010, ch. 3.
[38] K. Kikuchi, "Fundamentals of coherent optical fiber communications," Journal of lightwave technology, vol. 34, no. 1, pp. 157–179, Aug. 2015.
[39] G. P. Agrawal, "Fiber-optic communication systems," John Wiley & Sons; Feb. 2012, ch. 10.

[40] A. Awny, R. Nagulapalli, M. Kroh, J. Hoffmann, P. Runge, D. Micusik, G. Fischer, A. C. Ulusoy, M. Ko, D. Kissinger, "A linear differential transimpedance amplifier for 100-Gb/s integrated coherent optical fiber receivers," IEEE Transactions on Microwave Theory and Techniques, vol. 66, no. 2, pp. 973–986, Sep. 2017.
[41] S. Andreou, "Integrated stabilized laser systems for high resolution strain sensing," (Doctoral dissertation, Ph. D. thesis, Dept. Electr. Eng., Tech. Univ. Eindhoven, Eindhoven, The Netherlands), 2020, ch. 2.
[42] D. Pustakhod, K. Williams, X. Leijtens, "Fast and robust method for measuring semiconductor optical amplifier gain," IEEE Journal of Selected Topics in Quantum Electronics, vol. 24, no. 1, pp. 1-9, Aug. 2017.

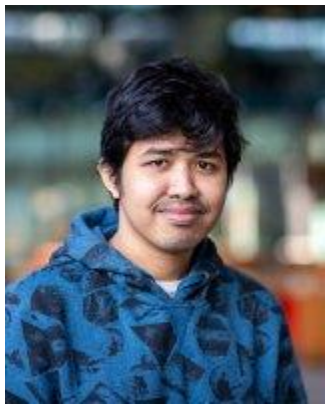

**Dhiman Nag** received his Bachelor of Engineering (B.E.) degree in Electronics and Communication from University Visvesvaraya College of Engineering, India in 2013. He received Master of Technology (M.Tech.) degree from R.V. College of Engineering, India in 2015. He was awarded with the Visvesvaraya fellowship to pursue Ph.D. in Indian Institute of Technology Bombay (IITB), India in 2016. He earned his Ph.D. degree in 2021 for his work in combating green gap in InGaN opto-electronics: from material to device perspective. He is currently pursuing his postdoc with the photonic integration group (PHI) in Eindhoven University of Technology (TUe), The Netherlands since 2021. He is involved in modelling and designing of coherent receiver from component to circuit level.

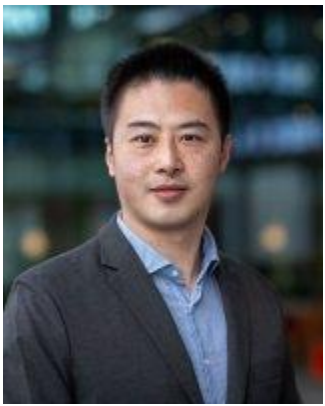

**Weiming Yao** (Member, IEEE) received the B.Sc. degree in electrical engineering from Technical University Berlin in 2010, the joint M.Sc. degree in photonic networks engineering from Aston University, U.K., and Scuola Superiore Sant'Anna, Italy, in 2012, and the Ph.D. degree for work on integrated high-capacity optical transceivers in 2017. Since 2012, he has been working with the Photonic Integration Group (PhI), Eindhoven University of Technology (TUe), The Netherlands. Since 2017, he has been with the Photonic Integration Technology Centre (PITC), where he has led an open innovation development line project for fabrication of photonic ICs (OpenPICs) and was co-applicant in the EU InPulse Pilot line project. Since 2020, he has been an Assistant Professor with TU/e. He was awarded a Dutch NWO Veni Grant in 2019 for research on integrated neuromorphic photonics.

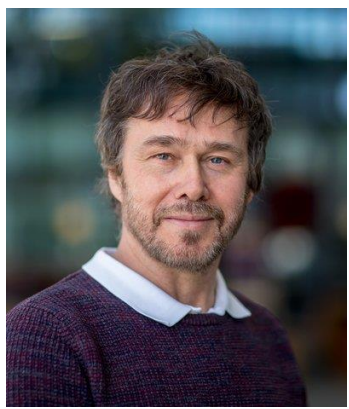

**Jos J. G. M. van der Tol** received the M.Sc. and Ph.D. degrees in physics from the State University of Leiden, The Netherlands, in 1979 and 1985, respectively. In 1985, he joined KPN Research, where he has been involved in the research on integrated optical components for use in telecommunication networks. Since July 1999, he was an Associate Professor with the Eindhoven University of Technology (TUe), The Netherlands, until his retirement at 2022, where his research interests include opto-electronic integration, polarization issues, photonic membranes, and photonic crystals. He is currently working as an emeritus professor in TUe. He is the (co)author of more than 250 publications in the fields of integrated optics and optical networks and has 25 patent applications to his name. His research interests include modeling of waveguides, the design of electro-optical devices on lithium niobate, and their fabrication. Furthermore, he has been working on guided wave components on III–V semiconductor materials. He has also been active in the field of optical networks, focusing on survivability, introduction scenarios, and management issues.